\documentclass[aps, reprint, groupedaddress, superscriptaddress, amsmath, amssymb]{revtex4-1}
\usepackage[colorlinks,allcolors=blue]{hyperref}
\usepackage{graphicx}
\usepackage{xcolor}
\usepackage{braket}

\begin{document}
\title{Charge Density Wave Hampers Exciton Condensation in 1T-TiSe$_2$}

\author{Chao Lian}
\affiliation{Department of Chemical \& Environmental Engineering, Materials Science \& Engineering Program, and Department of Physics \& Astronomy, University of California-Riverside, Riverside, CA 92521, USA.}
\author{Zulfikhar A. Ali}
\author{Bryan M. Wong}
\email{bryan.wong@ucr.edu}
\affiliation{Department of Chemical \& Environmental Engineering, Materials Science \& Engineering Program, and Department of Physics \& Astronomy, University of California-Riverside, Riverside, CA 92521, USA.}

\date{\today}

\begin{abstract}
The Bose-Einstein condensation of excitons continues to garner immense attention as a prototypical example for observing emergent properties from many-body quantum effects. In particular, Titanium Diselenide (TiSe$_2$) is a promising candidate for realizing exciton condensation and was experimentally observed only very recently [\href{https://doi.org/10.1126\%2Fscience.aam6432}{Science \textbf{358} 1314 (2017)}]. Surprisingly, the condensate was experimentally characterized by a soft plasmon mode that only exists near the transition temperature, $T_c$, of the charge density wave (CDW). Here, we characterize and analyze the experimental spectra using linear-response time-dependent density functional theory and find that the soft mode can be attributed to interband electronic transitions. At the CDW state below $T_c$, the periodic lattice distortions hamper the spontaneous formation of the exciton by introducing a CDW gap. The band gap raises the soft mode and merges it into the regular plasmon. Our surprising results contradict previous simplistic analytical models commonly used in the scientific literature. In addition, we find that a finite electronic temperature, $T_e$, introduces a dissipation channel and prevents the condensation above $T_c$. The combined effect of the CDW and $T_e$ explains the fragile temperature-dependence of the exciton condensation. Taken together, our work provides the first \textit{ab initio} atomic-level framework for rationalizing recent experiments and further manipulating exciton condensates in TiSe$_2$.
\end{abstract}

\maketitle

\section{Introduction.}
Bose-Einstein condensates (BECs) exhibit exotic transport phenomena such as superfluidity in liquid Helium and superconductivity via Cooper pairs. Other bosonic quasiparticles, such as excitons~\cite{Snoke2014, Seki2014, Snoke1990, Johnsen2001, Snoke1990a, Casella1963, Lin1993, Combescot2007, Kavoulakis2000, Cudazzo2010, Jiang2018d, Jiang2019, Butov2002, Blatt1962, Eisenstein2004, Mysyrowicz1990}, polaritons~\cite{Deng2002, Porras2002, Richard2005, Laussy2006, Wouters2007, Kasprzak2008, Utsunomiya2008, Lagoudakis2008, Deng2010, Ejaz2010}, and magnons~\cite{Ruegg2003, Radu2005, Demokritov2006, Popescu2011, Tserkovnyak2016} can also form a BEC, with exciton condensation particularly drawing immense recent attention~\cite{Kogar2017}. The exciton condensate is predicted to form a superfluid current~\cite{Hanamura1974,Haug1975}, which not only is an exotic emergent phenomenon in fundamental quantum research, but also vital for designing next-generation, scattering-free electronic devices.

\begin{figure*}
\centering
\includegraphics[width=1.0\linewidth]{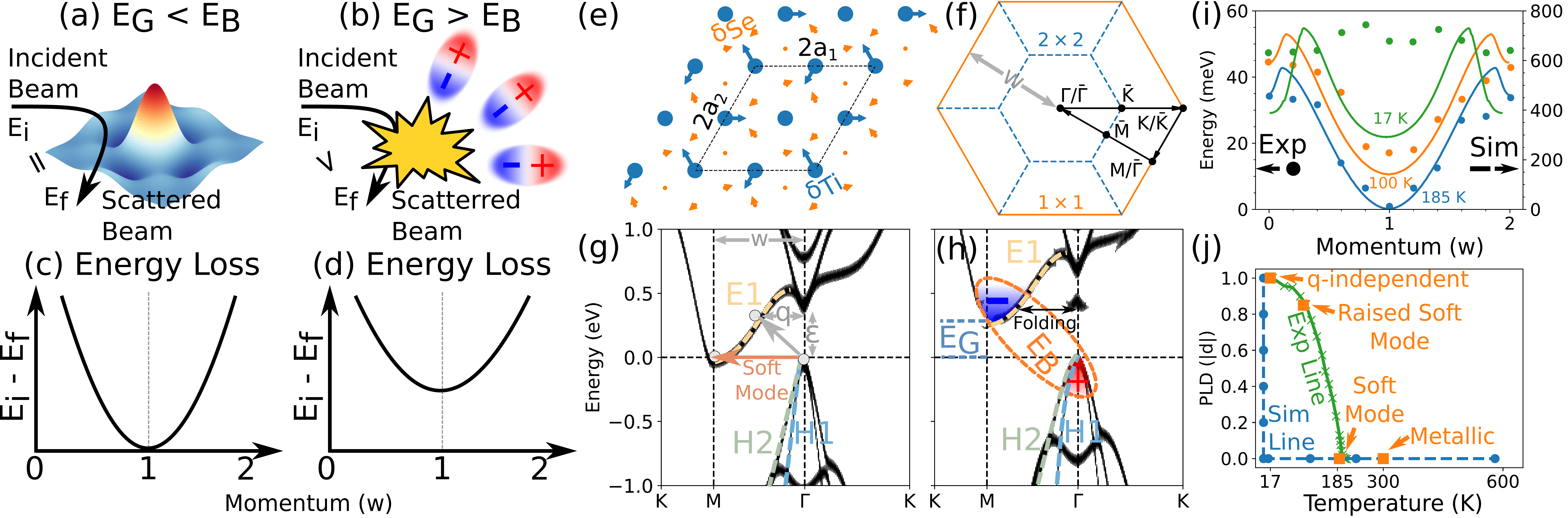}
\caption{(a-d) Schematic diagram of the exciton condensation. $E_G$ and $E_B$ are the band gap and excition binding energy, repectively, as shown in (h). $E_i$ and $E_f$ are the energies of the incident and scattered beam, respectively. $w$ is the momentum between the CBM and the VBM as shown in (g). (e) Atomic structure of 1T-TiSe$_2$. The blue and orange circles denote the Ti atoms and the Se atoms, respectively. The arrows denote the periodic lattice distortion (PLD) displacements \{$\mathbf{d}_i$\}. (f) The Brillouin zone (BZ) of TiSe$_2$. The solid orange and dashed blue lines denote the BZ of the $1\times1$ and $2\times2$ TiSe$_2$ cell, respectively. \{$\Gamma$, M, K\} and \{$\bar{\Gamma}$, $\bar{\mathrm{M}}$, $\bar{\mathrm{K}}$\} denote the special $k$ points in the $1\times1$ and $2\times2$ BZ, respectively. (g)(h) Band structures of 1T-TiSe$_2$ at (g) its normal state and (h) charge density wave state. (i) Experimental~\cite{Kogar2017} (dots) and simulated (solid lines) plasmon dispersions at different temperatures. (j) Phase diagram of the plasmon in TiSe$_2$ as a function of PLD and temperature. The blue circles and orange squares denote the experimental measurements and simulation results, respectively. The experimental line describes the PLD as a function of temperature, reproduced from~\cite{DiSalvo1976}.}
\label{fig:struct+BZ}
\end{figure*}

From a Bardeen-Cooper-Schrieffer (BCS)-like Hamiltonian, Kohn, J\'{e}rome, Halperin, and Rice proposed the phenomenon of exciton condensation in the 1960s~\cite{Kohn1967,Jerome1967, HALPERIN1968}. Simply put, for a material having an indirect gap, excitons spontaneously form if the exciton binding energy $E_B$ is larger than the band gap, $E_G$. When $E_G<E_B$, the total energy decreases by creating excitons with an identical momentum $q=w$; i.e., the exciton condensate, where $w$ is the reciprocal vector connecting the valence band maximum (VBM) and the conduction band minimum (CBM). The exciton condensate can be detected with momentum-resolved spectroscopy such as electron energy loss spectroscopy (EELS)~\cite{Egerton2009} and resonant inelastic X-ray scattering (RIXS)~\cite{Ament2011}. In these spectroscopic techniques, the incident electron or X-ray beam can induce a non-zero momentum excitation. As shown in Fig.~\ref{fig:struct+BZ}(a), when $E_G<E_B$, the exciton condensate elastically scatters the incident beam: the beam exchanges a momentum ${w}$ without energy loss~\cite{Kohn1967,Kogar2017}. This creates a soft mode in the interband plasmon at the momentum $q\to w$ [Fig.~\ref{fig:struct+BZ}(c)]. In comparison, when $E_G>E_B$, creating excitons consumes the energy of the incident beam [Fig.~\ref{fig:struct+BZ}(b)], thus the energy loss is always positive [Fig.~\ref{fig:struct+BZ}(d)]~\cite{Kohn1967,Kogar2017}.

Among the materials that meet this band structure requirement, TiSe$_2$ [Fig.~\ref{fig:struct+BZ}(e-h)] is a particularly promising candidate for observing exciton condensation effects~\cite{Monney2009}. Above the transition temperature $T_c \sim 190$~K, TiSe$_2$ has a negative indirect gap with the VBM at the M point and the CBM at the $\Gamma$ point, as shown in Fig.~\ref{fig:struct+BZ}(g). Besides its semimetallic nature, the quasi-two-dimensional (2D) structure of TiSe$_2$ weakens the Coulomb screening and favors exciton binding. In a pioneering experimental study, Kogar et al. recently observed exciton condensation in TiSe$_2$ using EELS~\cite{Kogar2017}. At $T_c$, the exciton condensation emerges as a soft plasmon mode near the edge of the Brillouin zone (BZ), which indicates that zero energy consumption is required to excite the electrons. 
In contrast to conventional BEC effects that are always stable at a sufficiently low temperature, the soft mode is fragile to both an increase and decrease of the temperature~\cite{Kogar2017}.  As summarized in Fig.~\ref{fig:struct+BZ}(i) and (j), instead of a sustainable soft plasmon mode, a momentum-independent plasmon and a regular metallic plasmon are observed at 17 and 300~K, respectively. The highly temperature-dependent exciton condensation reflects the rich physics in TiSe$_2$. Specifically, the well-known charge density wave (CDW) emerges below $T_c$, accompanied with a $2\times2$ periodic lattice distortion (PLD) \{$\mathbf{d}_i$\} shown in Fig.~\ref{fig:struct+BZ}(e). Although the CDW state has been extensively studied as a prototypical example of the CDW-superconductivity (SC) transition~\cite{Cui2006, Morosan2006, Li2007, Li2007b, Barath2008, Kusmartseva2009, Jeong2010, Zaberchik2010, Hillier2010, Giang2010, Morosan2010, Iavarone2012, Kacmarcik2013, Husanikova2013, Ganesh2014, Joe2014, Luna2015, Das2015, Medvecka2016, Wei2017b, Kogar2017a, Yan2017, Banerjee2018, Hildebrand2018, Yao2018}, the driving force of the CDW is still under debate regarding whether it is a pure electronic exciton-related mechanism~\cite{Rossnagel2002, Cercellier2007, Stoffel1982, Anderson1985a, Pillo2000, Kidd2002, Monney2010, Monney2010a, VanWezel2010, Monney2011, May2011, Monney2012, Monney2012a, Cazzaniga2012, Zenker2013, Koley2014, Watanabe2015, Peng2015b, Monney2015, Sugawara2016, Hildebrand2016a, Novello2017,Pasquier2018, Baranov2004, Chen2018c, Rohwer2011a} or an electron-phonon coupling (EPC) mechanism~\cite{Hughes1977, Wakabayashi1978, Gaby1981, Motizuki1981, Suzuki1984, Lopez-Castillo1987, Holt2001, Bussmann-Holder2009, VanWezel2010a, Rossnagel2010, Calandra2011, Weber2011, Zhu2012, Olevano2014, Wang2018, Kaneko2018}. The complex interplay among the thermal field, CDW order, and exciton condensation is beyond existing conventional models based on the assumption of fixed ions and single-particle bands~\cite{Kohn1967, Jerome1967, HALPERIN1968, Monney2009}. As such, a first-principles-based \textit{ab initio} framework is essential for understanding this entangled system and tuning the properties of the exciton condensation.

In this article, we accurately reproduce and characterize the experimental spectra using linear-response time-dependent density functional theory (lr-TDDFT). At the normal state, we observe a soft plasmon mode that directly represents the exciton condensation, which can be attributed to interband electronic transitions. At the CDW state below $T_c$, the periodic lattice distortions introduce a CDW gap and hamper the spontaneous formation of the exciton. Above $T_c$, the higher electronic temperature prevents the condensation by introducing a dissipation channel. The combined effect of the CDW and finite electronic temperature explains the fragile temperature-dependence of the exciton condensation. Our work provides the first \textit{ab initio} atomic-level framework, beyond widely-used simplistic analytical models, for rationalizing recent experiments and further manipulating exciton condensates in TiSe$_2$.
\section{\label{txt:RefbComc} Methodology} 
The ground state and lr-TDDFT calculations were carried out with the \textsf{GPAW}~\cite{Mortensen2005,Enkovaara2010,HjorthLarsen2017} package. The projector augmented-waves (PAW)~\cite{Blochl1994} method and Perdew-Burke-Ernzerhof (PBE) XC functional~\cite{Perdew1996} were used. The plane-wave cutoff energy was set to 750~eV. The on-site Coulomb interaction of the 3d orbitals of Ti was 3.5~eV. The Brillouin zone was sampled using the Monkhorst-Pack (MP) scheme~\cite{Monkhorst1976} with a $48\times48\times1$ k-point mesh. The computational cell examined in this work was a $2\times2$ supercell of TiSe$_2$, containing 4 Ti atoms and 12 Se atoms. In the lr-TDDFT calculations, the Bootstrap XC kernel~\cite{Sharma2011} was utilized together with the random phase approximation (RPA) as a comparison.

We calculate the EELS, i.e. the frequency and wave-vector dependent density response functions, based on the lr-TDDFT formalism~\cite{Yuan2009,Yan2011,Lian2018BoroPlasmon}. The non-interacting density response function in real space is written as
\begin{equation}
\begin{split}
\chi^0(\mathbf{r}, \mathbf{r}^{\prime}, \omega) =& \sum_{\mathbf{k}, \mathbf{q}}^{\mathrm{BZ}}\sum_{n, n^{\prime}}  
\frac{f_{n\mathbf{k}}-f_{n^{\prime} \mathbf{k} + \mathbf{q}}}{\omega + \epsilon_{n\mathbf{k}} - \epsilon_{n^{\prime} \mathbf{k} + \mathbf{q}} + i\eta} \times \\ 
& \psi_{n\mathbf{k}}^{\ast}(\mathbf{r}) \psi_{n^{\prime} \mathbf{k} + \mathbf{q} }(\mathbf{r}) \psi_{n\mathbf{k}}(\mathbf{r}^{\prime}) \psi^{\ast}_{n^{\prime} \mathbf{k} + \mathbf{q} }(\mathbf{r}^{\prime}),
\end{split}
\end{equation} 
where $n$ is the band index, $\mathbf{k}$ is the $k$ index, $\mathbf q$ stands for the Bloch vector of the incident wave, $\eta \rightarrow 0$, and $\epsilon_{n \mathbf{k}}$ and $\psi_{n \mathbf{k}}(\mathbf{r})$ are the eigenvalues and eigenvectors of the ground state Hamiltonian, respectively. 
The full interacting density response function is obtained by solving Dyson's equation from its non-interacting counterpart $\chi^0$ as
\begin{equation}
\begin{split}
\chi(\mathbf r, \mathbf{r^{\prime}}, \omega) =& \chi_0(\mathbf r,  \mathbf{r^{\prime}}, \omega) \\ 
+& \iint_{\Omega} d\mathbf{r}_1 d\mathbf{r}_2 \chi_0(\mathbf r, \mathbf{r}_1, \omega)
K(\mathbf{r}_1, \mathbf{r}_2) \chi(\mathbf{r}_2,  \mathbf{r^{\prime}} ,\omega),
\end{split}
\end{equation}
where the kernel is the summation of the coulomb and exchange-correlation (XC) interaction
\begin{equation}
K(\mathbf{r}_1, \mathbf{r}_2) = \frac{1}{|\mathbf{r}_1 -\mathbf{r}_2|} 
+ f_{xc}.
\end{equation}
Here, $f_{xc} = {\partial V_{xc}[n]}/{\partial n}$ is the XC kernel. The commonly-used XC kernels include the adiabatic local density approximation (ALDA)~\cite{Perdew1996}, Bootstrap approximation~\cite{Sharma2011}, etc. One of the simplest cases is the random phase approximation (RPA), with $f_{xc} = 0$.

For a system possessing translational symmetry, it is more convenient to represent $\chi^0$ in the reciprocal lattice space:
\begin{equation}
\chi^0(\mathbf{r}, \mathbf{r}^{\prime},  \omega) = \frac{1}{\Omega} 
\sum_{\mathbf{q}}^{\mathrm{BZ}} \sum_{\mathbf{G} \mathbf{G}^{\prime}}
e^{i(\mathbf{q} + \mathbf{G}) \cdot \mathbf{r}} \chi^0_{\mathbf{G} \mathbf{G}^{\prime}}(\mathbf{q}, \omega) 
e^{-i(\mathbf{q} + \mathbf{G}^{\prime}) \cdot \mathbf{r}^{\prime}} ,
\end{equation} 
where $\Omega$ is the normalization volume and $\mathbf G (\mathbf G^{\prime})$ are reciprocal lattice vectors.
The Fourier coefficients $\chi^0_{\mathbf{G} \mathbf{G}^{\prime}}(\mathbf{q}, \omega)$ are written as
\begin{equation}
\label{eq:screening}
\begin{split}
\chi^0_{\mathbf{G} \mathbf{G}^{\prime}}(\mathbf{q}, \omega) = & 
\sum_{n, n^{\prime}} \chi^0_{\mathbf{G} \mathbf{G}^{\prime} n, n^{\prime} }(\mathbf{q}, \omega)
\end{split}
\end{equation}
where
\begin{equation}
\label{chi0nn'}
\begin{split}
\chi^0_{\mathbf{G} \mathbf{G}^{\prime} n, n^{\prime} }(\mathbf{q}, \omega) = & \frac{1}{\Omega} 
\sum_{\mathbf{k}}^{\mathrm{BZ}} 
\frac{f_{n\mathbf{k}}-f_{n^{\prime} \mathbf{k} + \mathbf{q} }}{\omega + \epsilon_{n\mathbf{k}} - \epsilon_{n^{\prime} \mathbf{k} + \mathbf{q} } + i\eta} \\ 
&\times \langle \psi_{n \mathbf{k}} | e^{-i(\mathbf{q} + \mathbf{G}) \cdot \mathbf{r}} | \psi_{n^{\prime} \mathbf{k} + \mathbf{q} } \rangle_{\Omega_{\mathrm{cell}}} \\
&\times \langle \psi_{n\mathbf{k}} | e^{i(\mathbf{q} + \mathbf{G}^{\prime}) \cdot \mathbf{r}^{\prime}} | \psi_{n^{\prime} \mathbf{k} + \mathbf{q} } \rangle_{\Omega_{\mathrm{cell}}},
\end{split}
\end{equation}
Dyson's equation is expressed in the $\mathbf{G}$ basis as
\begin{equation}
\label{eq:Dyson}
\begin{split}
\chi_{\mathbf G \mathbf G^{\prime}}(\mathbf q, \omega)  
=& \chi^0_{\mathbf G \mathbf G^{\prime}}(\mathbf q, \omega) \\
+& \sum_{\mathbf G_1 \mathbf G_2} \chi^0_{\mathbf G \mathbf G_1}(\mathbf q,  \omega) K_{\mathbf G_1 \mathbf G_2}(\mathbf q)
\chi_{\mathbf G_2 \mathbf G^{\prime}}(\mathbf q, \omega). 
\end{split}
\end{equation}
The dielectric function can be expressed with $\chi_{\mathbf G \mathbf G^{\prime}}(\mathbf q, \omega)$ as
\begin{equation}
\epsilon^{-1}_{\mathbf G \mathbf G^{\prime}}(\mathbf q, \omega)
= \delta_{\mathbf G \mathbf G^{\prime}} - \sum_{\mathbf{G}_1} K_{\mathbf{G} \mathbf{G}_1}(\mathbf q) \chi_{\mathbf{G}_1 \mathbf G^{\prime}}(\mathbf q, \omega).
\end{equation}
The macroscopic dielectric function is defined by
\begin{equation}
\epsilon_M(\mathbf q, \omega) = \frac{1}{\epsilon^{-1}_{00}(\mathbf q, \omega)},
\end{equation}
and the electron energy loss spectrum (EELS) is 
\begin{equation}
\label{eq:EELS}
\mathrm{EELS} = -\mathrm{Im}\frac{1}{\epsilon_M(\mathbf q, \omega)}.
\end{equation}

\begin{figure}
\centering
\includegraphics[width=1\linewidth]{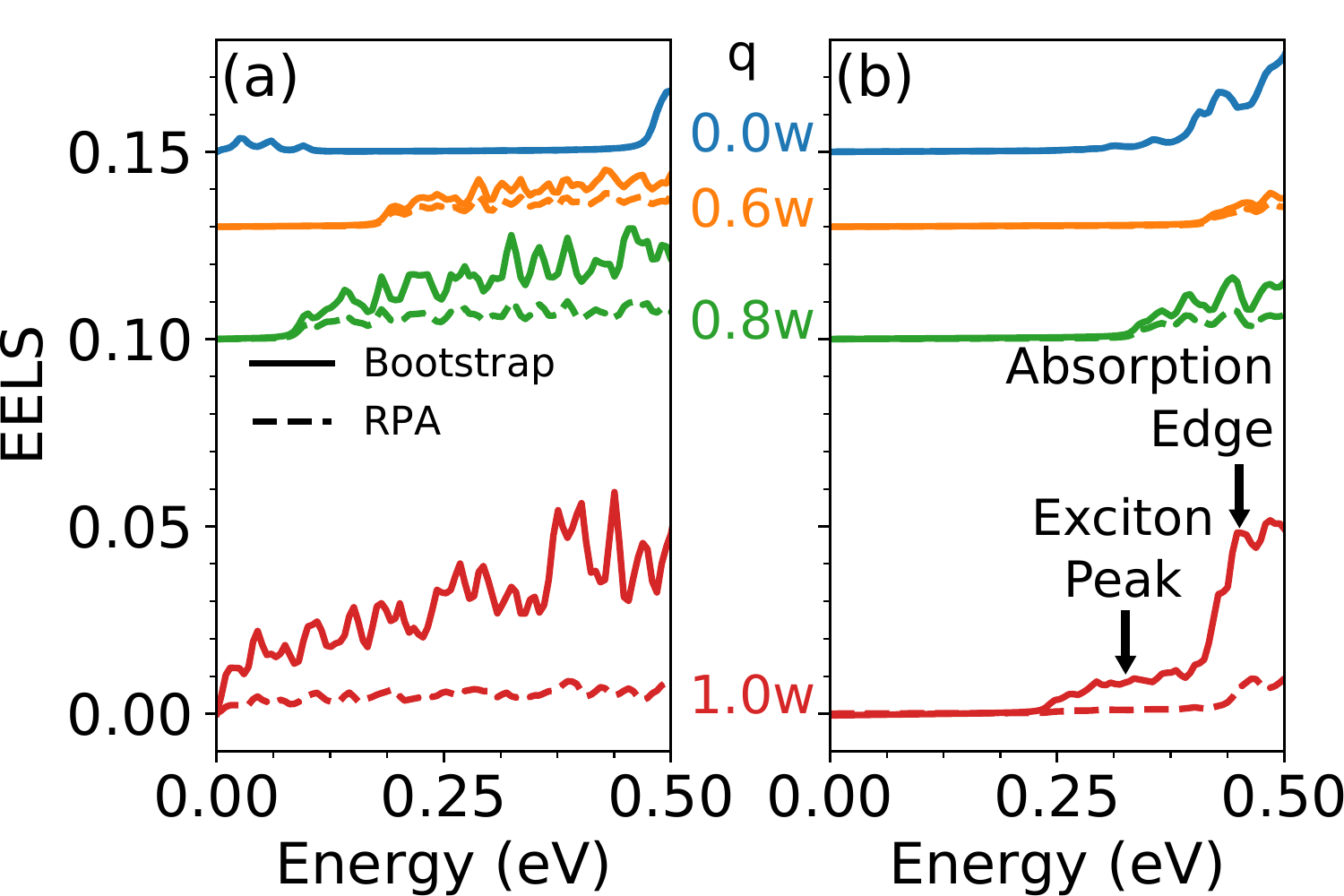}
\caption{Comparison of the EELS with the random phase approximation (RPA) vs Bootstrap XC kernels for (a) the normal state (b) the CDW state.}
\label{fig:RPAvsBoot}
\end{figure}
We briefly discuss the accuracy of lr-TDDFT in describing the exciton in TiSe$_2$. It is known that semi-local XCs (e.g. PBE) poorly describe long-range Coulomb screening~\cite{Sharma2011}, and long-range corrections, such as in the Bethe–Salpeter equation (BSE) and long-range-corrected XC functionals~\cite{doi:10.1063/1.2907858,doi:10.1021/acs.jctc.7b00423,B901743G,doi:10.1021/ct100727g,doi:10.1021/jp9074674,doi:10.1021/acs.jctc.5b00105}, are required for improved accuracy. However, the exciton in TiSe$_2$ is formed by an attractive interaction $V(w)$ between the electron pocket at the M point and the hole pocket at the $\Gamma$ point, as shown in Fig.~\ref{fig:struct+BZ}(d). Here, $w=\pm 0.5|\mathbf{b}_1|$ is the reciprocal vector and $\mathbf{b}_1$ ($i=1,2$) is the reciprocal lattice vector along the $i$th direction, as shown in Fig.~\ref{fig:struct+BZ}(c). Thus, $V(w)$ is a medium-ranged interaction with a characteristic distance of $1/w = a$, where $a=|\mathbf{a_i}|$ and $\mathbf{a}_i$ is the lattice vector. $V(w)$ is distinctly different from typical long-range interactions that are characteristic of vertical excitonic excitations in momentum space.

We quantitatively characterize the exciton effect by comparing the EELS calculated with the RPA and the Bootstrap XC kernel, as shown in Fig.~\ref{fig:RPAvsBoot}. The Bootstrap XC kernel is designed to correct the long-range error of the Coulomb interaction and generates accurate exciton peaks at a similar computational cost of ALDA~\cite{Sharma2011}. The bootstrap XC kernel can yield a good description of exciton when screening is not high that is suitable for many low-dimensional and semi-metallic systems. At $q \sim 0$, the EELS spectra with the RPA and Bootstrap are generally the same, indicating the absence of the exciton-type excitation. We note that with increasing $q$, the difference increases and reaches a maximum at $q=w$, corresponding to the emergence of exciton excitations. As shown in Fig.~\ref{fig:RPAvsBoot}(b), a new exciton peak can be observed only in the Bootstrap calculation at $q \sim 0$ around $320$~meV, which is $100$~meV lower than the absorption edge. This indicates that the exciton binding energy $E_B$ in the Bootstrap calculations is consistent with the results from BSE, $E_B=75$~meV~\cite{Pasquier2018}.

The CDW phase of TiSe$_2$ is a $2\times2$ cell of the normal phase. The energy bands are folded from the $1\times1$ BZ to the $2\times2$ BZ. In contrast, ARPES measurements still span over the $1\times1$ BZ. To bridge the gap between the DFT bands and the measured ARPES spectra, the band unfolding technique is used to calculate the effective band structure (EBS)~\cite{Ku2010,Popescu2012} of the supercell (SC). Expanding the adiabatic basis $\ket{\phi_{i,\mathbf{k}}}$ of the $2\times2$ SC in the wavefunction $\ket{\Phi_{I,\mathbf{K}}}$ of primitive $1\times1$ cell (PC), we get 
\begin{equation}
\ket{\phi_{i,\mathbf{k}}(\mathbf{G},t)}=\sum_{I,\mathbf{K}}a(I,\mathbf{K};i,\mathbf{k};t)\ket{\Phi_{I,\mathbf{K}}\mathbf{G},t)},
\end{equation}
where $\mathbf{K}=\mathbf{k}+\mathbf{B}$, $\mathbf{B}$ is the reciprocal basis vector of SC, and $a(I,\mathbf{K};i,\mathbf{k};t)$ is the coefficient of $\ket{\Phi_{I,\mathbf{K}}}$ as the basis of $\ket{\phi_{i,\mathbf{k}}}$. The spectral function is the EBS along the $\mathbf{K}$ path in PCBZ:
\begin{equation}
A(\mathbf{K}, E, t)=\sum_i P(\mathbf{K};\mathbf{k},i;t)\delta(E-\epsilon_{i,\mathbf{k}}),
\end{equation}
where $\epsilon_{i,\mathbf{k}}$ is the eigenvalue of band $i$ at the $\mathbf{k}$ point, $E$ is the energy, and 
\begin{equation}
\label{probability}
P(\mathbf{K};\mathbf{k},i) =\sum_{\mathbf{G}}\left|\phi_{i,\mathbf{k}}(\mathbf{G}+\mathbf{K}-\mathbf{k}, t)\right|^2,
\end{equation}
We can introduce an extra weight function $w(i,\mathbf{k})$ 
\begin{equation}
\label{eq:spacial_band_unfolding}
A(\mathbf{K},E,t)=\sum_i P(\mathbf{K};\mathbf{k},i;t) w_{i,\mathbf{k}} \delta(E-\epsilon_{i,\mathbf{k}}).
\end{equation} 
The choice of $w_{i,\mathbf{k}}$ is arbitrary~\cite{Lian2017SiUnfold}. Here, we use the Fermi-Dirac distribution 
\begin{equation}
\label{eq:weight_function}
w_{i,\mathbf{k}} = \left[1 + \exp\left(\frac{\epsilon_{i,\mathbf{k}}-E_F}{T_e}\right)\right]^{-1},
\end{equation}
where $E_F$ is the Fermi energy and $T_e$ is the electronic temperature.

\section{Results and Discussion}
We calculate the structures of the normal 1$\times$1 and 2$\times$2 CDW phase and obtain the optimized PLD displacements \{$\mathbf{d}_i$\} shown in Fig.~\ref{fig:struct+BZ}(e), with $\delta_{Ti} = 0.083$~\AA, $\delta_{Se} = 0.027$~\AA, and $\delta_{Ti}/\delta_{Se} = 3.07:1$. These results accurately reproduce the experimental measurements of $\delta_{Ti}=0.085 \pm 0.014$~\AA\ and $\delta_{Ti}/\delta_{Se}\sim3:1$~\cite{DiSalvo1976}. To directly compare with the angle-resolved photoemission spectroscopy (ARPES) measurements, we unfold the energy bands from the $2\times2$ BZ to the $1\times1$ BZ to generate the effective band structures (EBS) along the K-M-$\mathrm{\Gamma}$-K symmetry points, as shown in Fig.~\ref{fig:struct+BZ}(f). In its normal state, TiSe$_2$ is semimetallic [Fig.~\ref{fig:struct+BZ}(g)]: the conduction band E1 touches the Fermi energy at the M point; two valence bands, H1 and H2, touch the Fermi energy at the $\Gamma$ point. At low temperatures, the CDW state of 1T-TiSe$_2$ is a semiconductor. The $2\times2$ CDW order backfolds the H1 band to the M point. Induced by the avoided crossing, a repulsive force raises the E1 band and opens a gap of $E_G \sim 0.27$~eV [Fig.~\ref{fig:struct+BZ}(h)]. Via the same mechanism, the backfolded E1 band at the $\Gamma$ point cuts the H1 and H2 bands into valence and conduction parts and upshifts the conduction parts of the H1 and H2 bands.

\begin{figure*}
\centering
\includegraphics[width=1\linewidth]{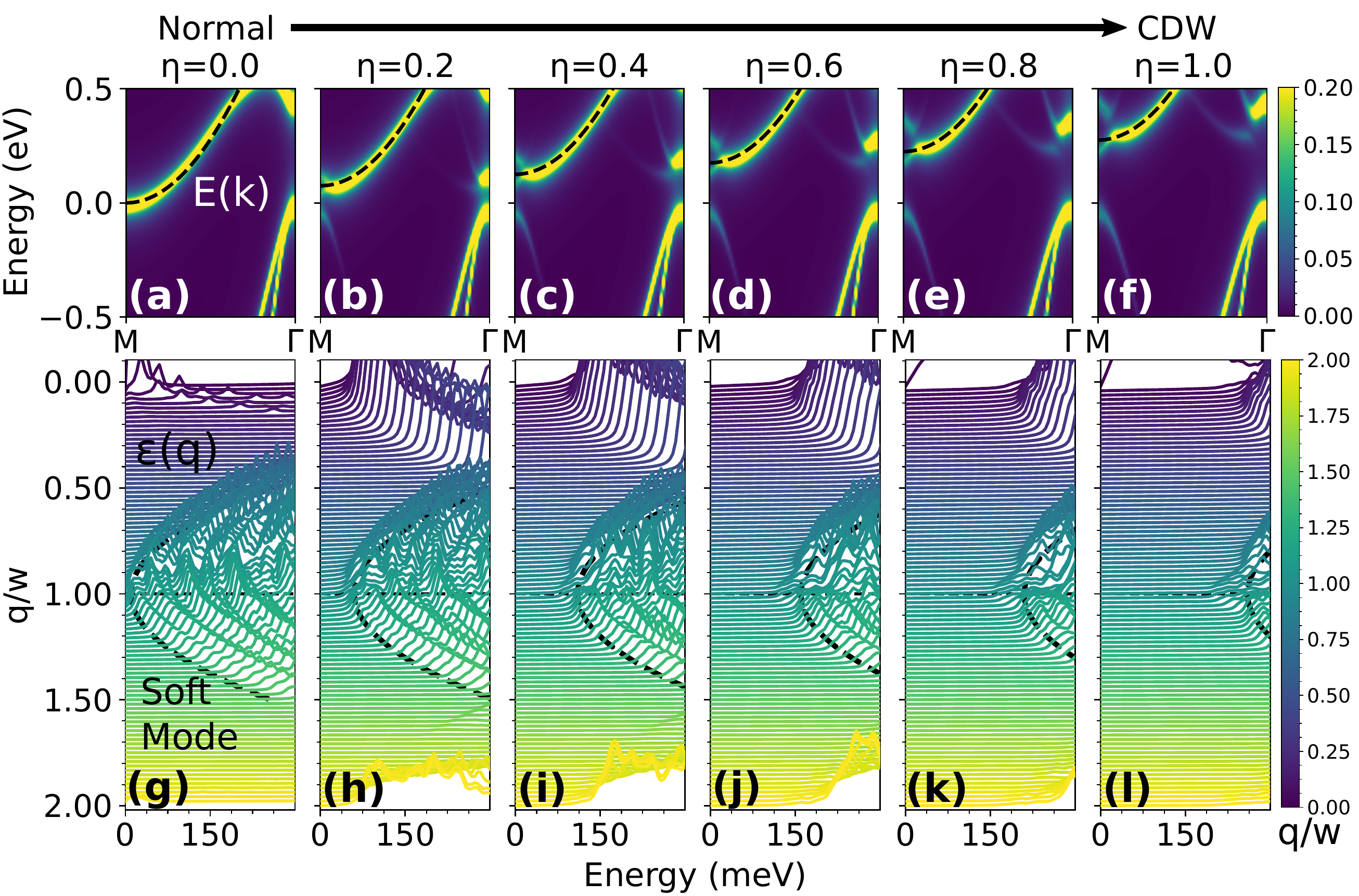}
\caption{(a-f) Effective band structures (EBS) and (g-l) electron energy loss spectra (EELS) at different PLDs \{$\eta \mathbf{d}_i$\}. The colorbars for (a-f) and (g-l) denote the spectral weight [Eq.~\ref{eq:spacial_band_unfolding}] of the EBS and the momentum $q$ of the EELS [Eq.~\ref{eq:EELS}], respectively. The dashed lines in panels (a-f) and (g-l) denote the parabolic fittings of the band structures $E(k)$ [Eq.~\ref{eq:bandE1}] near the M point and the plasmon dispersions $\epsilon(q)$ [Eq.~\ref{eq:softmode}] near $w$, respectively.}
\label{fig:PLDvD}
\end{figure*}

As previously mentioned, the soft mode is highly temperature-dependent and distinguishable from other BEC phenomena. Thus, the knowledge of thermal field effects is essential for understanding the exciton condensation. We propose that the thermal field mainly influences the electronic structures in two aspects:
\textbf{(I)} The decrease in temperature stabilizes the PLD and introduces a CDW gap. 
\textbf{(II)} The increase in temperature of the electron system, $T_e$, creates more thermal carriers. 
As shown in Fig.~\ref{fig:struct+BZ}(j), both $T_e$ and PLD cannot be dis-entangled in the experiment, and only the states on the experimental line can be observed. Nevertheless, our \textit{ab initio} techniques are not limited to these experimentally-observed states and are capable of further sampling the phase diagram and characterizing the effects of the PLD and $T_e$ separately.

We first examine the effects of the PLD by calculating the EBS and EELS for 6 structures with different PLDs \{$\eta \mathbf{d}_i$\} at $T_e =0$~K, where $\eta =  0.0, 0.2, ..., 1.0$, i.e., the states along the Y axis in Fig~\ref{fig:struct+BZ}(j). $\eta = 0.0$ and $\eta = 1.0$ correspond to the normal and CDW state, respectively. The intermediate $\eta$ values describe the structures at low but finite temperatures $0<T<T_c$.

We observed the soft mode [Fig~\ref{fig:struct+BZ}(g)] at the normal state with $\eta=0$. As shown in Fig.~\ref{fig:PLDvD}(a) and (g), the interband plasmon energy $\epsilon(q)$ decreases with $q<w$, reaches the zero point $\epsilon(q)=0$~eV at $q=w$ and then increases with $q>w$. This behavior is consistent with the experimental observations~\cite{Kogar2017}, although the rate of change is higher in our simulation. The experimental energy range of the soft mode is 0--40~meV, whereas we obtain a range of 0--400~meV from our simulations, which is due to the underestimation of the Coulomb screening in the semi-local exchange-correlation functionals. 

The CDW gap increases with $\eta$, as shown in Fig.~\ref{fig:PLDvD}(a-f). To quantitatively describe the effect of the PLD, we construct an expression for this band-plasmon correspondence. As shown in Fig.~\ref{fig:PLDvD}(a-f), the low-energy-range band can be accurately described by the expression 
\begin{equation}
\label{eq:bandE1}
E(k) = \max\{\alpha (k-k_M)^2 + E_G, 0\},
\end{equation}
where $k_M$ is the $k$ coordinates of the M point, $\alpha = 0.95$~eV/\AA$^2$, and $E_G=250\eta$~meV are the CDW gaps. As shown in Fig.~\ref{fig:PLDvD}(g-l), we find that the interband plasmon dispersion near $w$ can be accurately described by the expression
\begin{equation}
\label{eq:softmode}
\epsilon(q) = \max\{\alpha (q-w)^2 + E_G, 0\},
\end{equation}
with the same values of $\alpha = 0.95$~eV/\AA$^2$ and $E_G=250\eta$~meV in Eq.~\ref{eq:bandE1}. This indicates that the incident beam pumps electrons from $\Gamma$ to the E1 band, producing the EELS signal near $w$, as shown in Fig.~\ref{fig:struct+BZ}(g). At the normal state, we observed a soft mode with zero excitation energy at $q=w$. In addition, the band gap increases with $\eta$ and consequently raises the soft mode to a higher energy. The excitation energy at $q=w$ concomitantly increases from zero to a finite value, preventing the spontaneous exciton condensation.

\begin{figure*}
\centering
\includegraphics[width=1\linewidth]{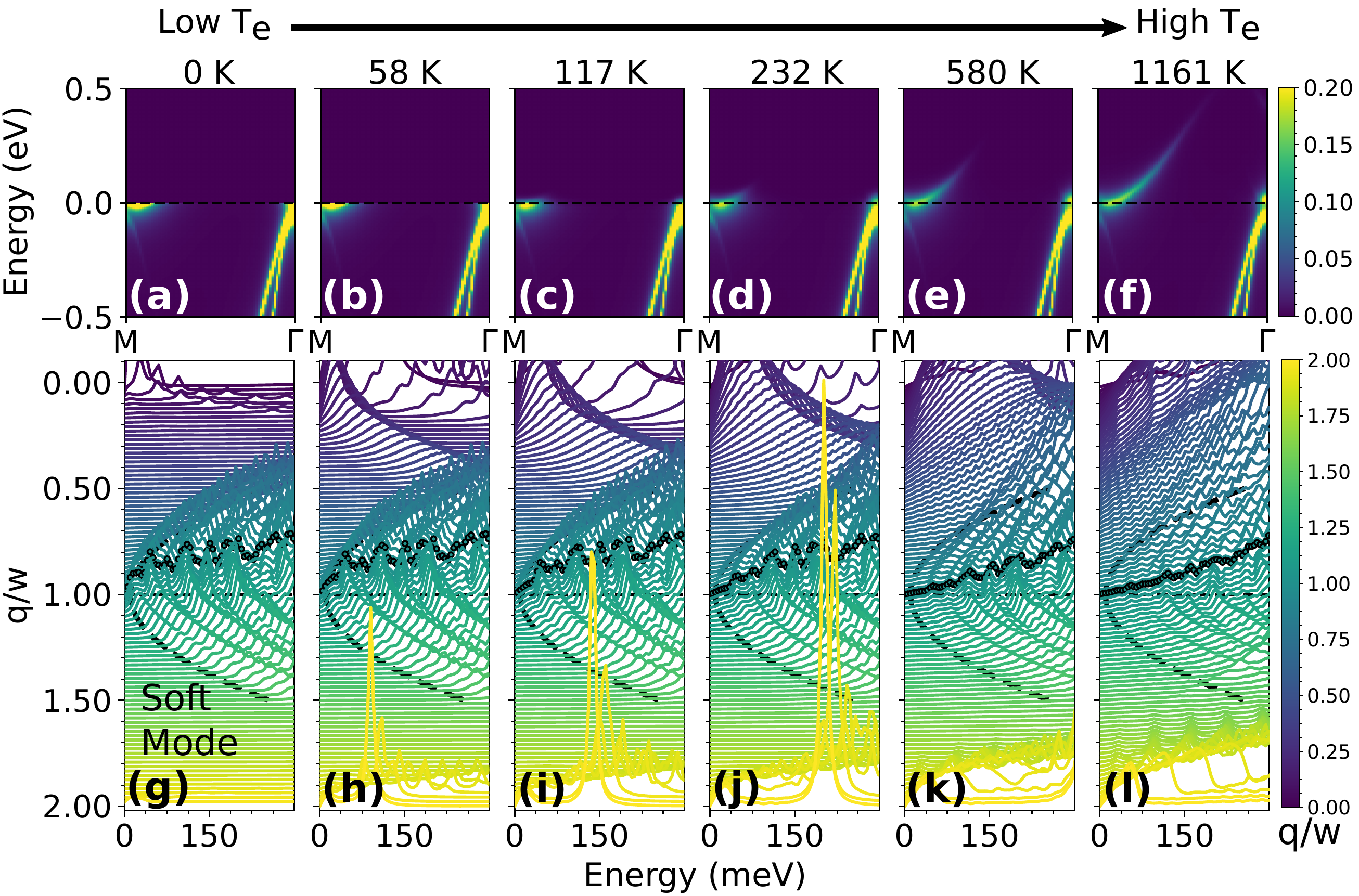}
\caption{(a-f) Effective band structures (EBS) and (g-l) electron energy loss spectra (EELS) at different electronic temperatures $T_e$. The colorbars for (a-f) and (g-l) denote the electron population of EBS [Eq.~\ref{eq:spacial_band_unfolding} and~\ref{eq:weight_function}] and the momentum $q$ of the EELS [Eq.~\ref{eq:EELS}], respectively. The dashed lines in (g)-(l) denote the plasmon dispersions [Eq.~\ref{eq:softmode}] near $w$. The dark dots denote the EELS at $q=w$.}
\label{fig:PLDvT}
\end{figure*}

Next, we discuss the influence of electronic temperature $T_e$ on the exciton condensation. The finite electronic temperature $T_e$ is approximated by broadening the Fermi-Dirac distribution as
\begin{equation}
f(E) = \frac{1}{1+\exp[\left(E-E_F\right)/k_B T_e]},
\end{equation}
where $E$ is the energy of the electron, $E_F$ is the Fermi energy, and $k_B$ is the Boltzmann constant. Due to the semi-metallic nature of TiSe$_2$, the carrier density is fairly sensitive to the temperature. We simulated the EELS at a different electronic temperature $T_e$ with $\eta=0.0$, as shown in Fig.~\ref{fig:PLDvT}. The increase in $T_e$ changes the electron distribution but barely alters the band structure [Fig.~\ref{fig:PLDvT}(a)-(f)]. Accordingly, the thermal carriers gradually fill the empty bands as $T_e$ increases. Compared with the unified quantum state of the exciton condensate, the thermal carriers form excitons with a wide range of momenta. The thermally-excited excitons with random momenta collide with and accelerate the decoherence of the exciton condensate. As shown in Fig.~\ref{fig:PLDvT}(g)-(l), the intensity of the exciton excitation decreases due to the partially-occupied initial and final state. This effect broadens the peak and decreases the peak height without changing the peak positions of the soft mode.

Besides the dissipation introduced by the thermal charge carriers, the increasing phonon density may also destroy the exciton condensation. Both linear-response time-dependent density functional theory (lr-TDDFT)~\cite{Yuan2009,Yan2011,Lian2018BoroPlasmon} and density-functional perturbation theory (DFPT)~\cite{Baroni2001, Giustino2017} are required to describe the phonon-induced linewidth broadening in the plasmon spectra. To the best of our knowledge, although temperature-dependent quasiparticle linewidths and optical spectra can be calculated with several DFT packages such as \textsf{Abinit}~\cite{Ponce2014}, \textsf{Yambo}~\cite{Ponce2015}, and \textsf{EPW}~\cite{Ponce2016}, \textbf{k}-resolved algorithms to calculate temperature-dependent plasmon spectra are still under development. Thus, we only considered the dissipation induced by thermal charge carriers. A large density of acoustic phonons will definitely accelerate the dissipation of exciton condensation at high temperatures, while its mechanism is still an important open question.

We note that frequency-independent xc kernels such as the adiabatic local density approximation and bootstrap~\cite{Sharma2011} neglect disorder and phonon scattering. In addition, the damping due to electronic many-body effects is absent. This effect can be introduced in the non-adiabatic xc kernel based on the time-dependent current density functional formalism~\cite{Vignale1996,Wijewardane2005,Ullrich2002,Ullrich2004a,Ullrich2006,Lacombe2019}, which shows good quantitative agreement with experimental linewidth data~\cite{Ullrich2004,Vignale1997,Ullrich2001,Ullrich1998,Berger2015}. Furthermore, increasing the temperature introduces extra damping from the enhanced phonon scattering which will lead to a faster vanishing of the soft mode signals when the temperature increases above $T_c$.


\begin{figure}
\centering
\includegraphics[width=1.0\linewidth]{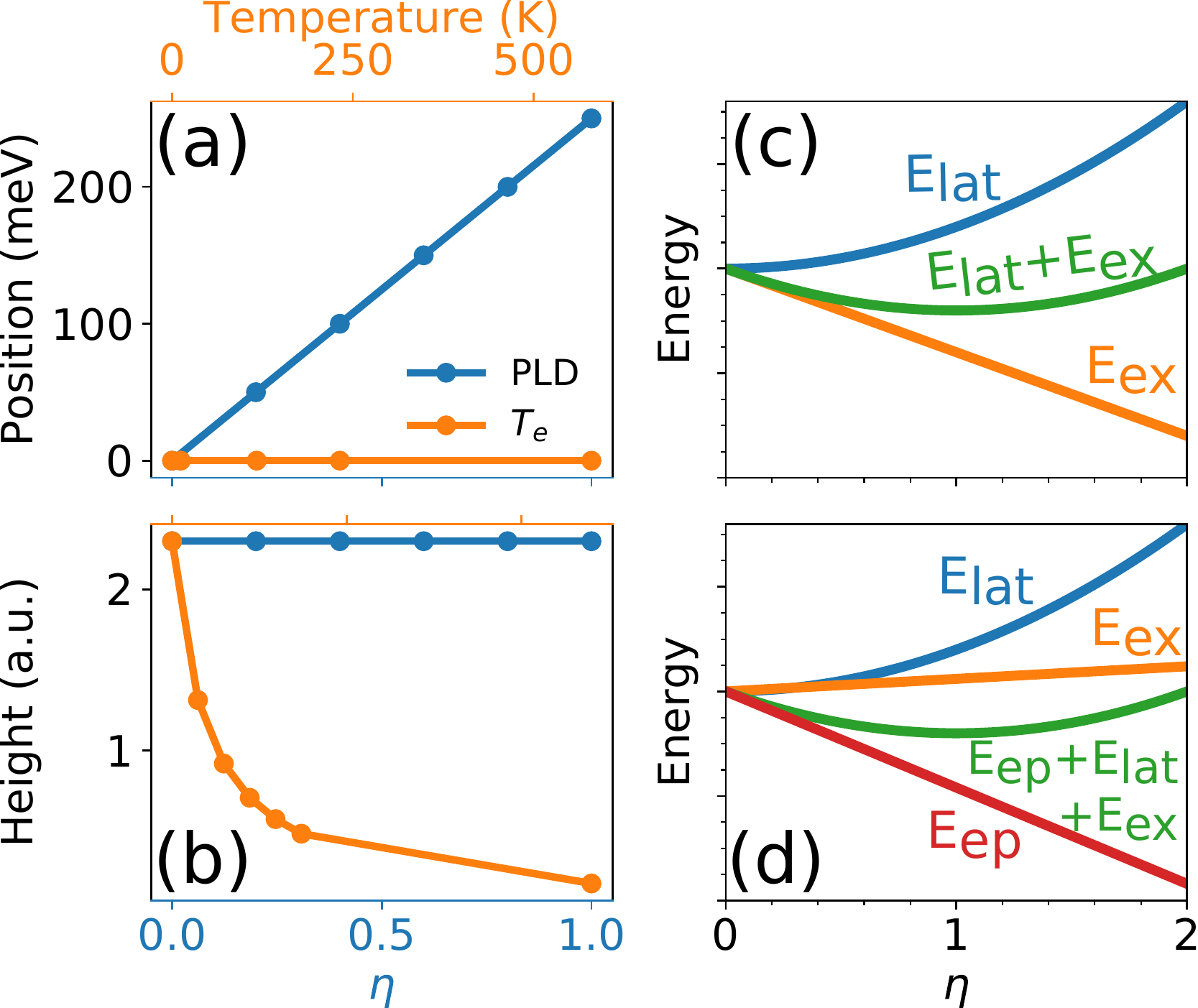}
\caption{The peak position $\epsilon(q=w)$ [Eq.~\ref{eq:softmode}] (a) and the peak height at $\epsilon(q=w)$ (b) of the soft mode as a combined effect of the PLD and $T_e$. (c)(d) The lattice energy $E_{\mathrm{lat}}$, exciton energy $E_\mathrm{ex}$, electron-phonon energy $E_\mathrm{ph}$, and total energy $E_\mathrm{tot}$ as a function of the PLD, indicated by (c) the analytical model and (d) our DFT and experimental framework.
}
\label{fig:Diagram}
\end{figure}

{Combining our analysis of the effects of the PLD and $T_e$, we have developed a microscopic framework to explain the temperature dependence of the plasmon dispersion. As shown in Fig.~\ref{fig:Diagram}(a) and (b), the PLD raises the soft mode without smearing the plasmon peaks; in contrast, the electronic temperature smears the peaks without changing the dispersion of the soft mode. Based on this framework, the experimental observations can be explained by recognizing that at a high temperature of $T=300$~K, TiSe$_2$ is semimetallic without a PLD. However, the thermal carriers occupy the E1 band and completely smear the soft mode signals and, therefore, only the regular metallic plasmon can be observed in the experimental EELS. At a low temperature of $T=17$~K, the thermal excitation is suppressed, while the energy required to create an exciton significantly increases due to the emerging CDW gap. When the transition temperature is near $T=185$~K, both the PLD and carrier density are sufficiently small, resulting in the observed soft modes.}

{To further verify our framework more quantitative, we directly compared calculated and experimental plasmon dispersions $\epsilon(T)$ at temperature $T$. The effect of temperature is simulated with \{$T_e=T$, $\eta=\eta(T)$\}, where $\eta=\eta(T)$ is the PLD as a function of temperature, reproduced from~\cite{DiSalvo1976} and shown in Fig.~\ref{fig:struct+BZ}(j). We compare the experimental plasmon dispersion at $T=17$~K, $T=100$~K, and $T=185~K$ with our calculations with the parameters \{$T_e=0$~K, $\eta=1.0$\}, \{$T_e=117$~K, $\eta=0.8$\}, and \{$T_e=232$~K, $\eta=0.0$\}, respectively. As shown in Fig.~\ref{fig:struct+BZ}(i), despite the difference in the absolute energy, the plasmon dispersions in our simulations are consistent with the experimental measurements for all three cases: at $T=17$~K, the CDW gap raises the interband plasmon mode to the same energy range of the regular plasmon, resulting in a momentum-independent plasmon dispersion; at $T=100$~K, the CDW gap decreases together with the PLD, presenting a soft mode with a finite energy gap at $q=w$; at $T_c=185$~K, the soft mode is observed in both the experimental and simulated EELS as a result of the absent CDW gap and relatively small thermal excitation.} 

The complex effects of the thermal field obtained from our \textit{ab initio} calculations provide a new mechanistic explanation that is quite distinct from the results of the analytical model commonly used in the scientific literature. In the analytical model, an additional BCS-like electron-hole coupling term~\cite{VanWezel2010a} is added to the single particle tight-binding Hamiltonian~\cite{Kohn1967, Jerome1967, HALPERIN1968, Monney2009}.
The exciton effect is overestimated because the BCS-like term is parameter-dependent, and the electron-phonon coupling is absent in the Hamiltonian. The single-particle band structures are fixed to those at the normal state, and the CDW gap is omitted when the temperature decreases to $T_c$. In Ref.~\cite{Monney2009}, the Hamiltonian without exciton interaction is
\begin{equation}
\label{Eq:tbham}
H_0 = \sum_{\mathbf{k}} \epsilon_v(\mathbf{k}) a^{\dagger}(\mathbf{k})a(\mathbf{k}) + \sum_{i,\mathbf{k}} \epsilon^i_c(\mathbf{k}) b_i^{\dagger}(\mathbf{k})b_i(\mathbf{k})
\end{equation}
Here $a^{\dagger}$ and $b^{\dagger}$ are operators creating electrons with wave vector $\mathbf{k}$ in the valence band and in the conduction band labeled $i$, respectively. To describe the exciton dissipation by the CDW, we propose a simple modification to the tight-binding model by adding the temperature-dependent CDW gap
\begin{equation}
\epsilon_{\mathrm{CDW}}(T) = \begin{cases}
0,& \text{if } T\geq T_c\\
(1-T/T_c)E_g,              & \text{if } T < T_c
\end{cases}
\end{equation}
to Eq.~\ref{Eq:tbham} as
\begin{equation}
H_0 = \sum_{\mathbf{k}} \epsilon_v(\mathbf{k}) a^{\dagger}(\mathbf{k})a(\mathbf{k}) + \sum_{i,\mathbf{k}} \left[\epsilon^i_c(\mathbf{k}) b_i^{\dagger}(\mathbf{k})b_i(\mathbf{k}) + \epsilon_{\mathrm{CDW}}(T)\right]
\end{equation}
where $E_g = 0.27$~eV is the CDW gap at $T=0$ and $T_c = 180$~K is the transition temperature.
As shown in Fig.~\ref{fig:Diagram}(c), the analytical model indicates that the CDW state is a compromise between the increasing lattice energy and the decreasing electron energy via the formation of excitons~\cite{VanWezel2010a}. This indicates that the exciton condensation distorts the lattice and lowers the energy of the system. Thus, the soft mode should not be destroyed by the PLD, provided that $T<T_c$. Instead, both the experimental spectra~\cite{Kogar2017} and our EELS simulation show a different phenomenon: the PLD hinders the exciton condensation, which indicates another mechanism such as EPC is essential for stabilizing the CDW state [Fig.~\ref{fig:Diagram}(d)].

We note that the mechanism presented for the low-temperature suppression of condensation does not depend on the existence of a CDW phase. The isolated electron system may form an exciton condensation if the Coulomb binding is larger than the bandgap~\cite{Kohn1967,Jerome1967, HALPERIN1968}. However, the interactions between the electron and phonon in this system introduces a channel for energy redistribution; specifically, the ionic energy can be increased by forming the PLD, and the electronic energy can be decreased by opening a CDW gap, which lowers the total energy. Additional mechanisms may also raise the energy of the soft plasmon mode in materials where no such CDW phase exists. We expect that interactions between electrons and different systems (other than phonons) may also introduce a channel for energy redistribution. For example, the interaction between the electron and spin may introduce a periodic spin order; i.e., a spin density wave (SDW)~\cite{Grner1994}. This will also open a bandgap and suppress the exciton condensation. An SDW state instead of the exciton condensation state emerges under the transition temperature.

\section{Conclusions.}
In summary, we have developed a new \textit{ab initio} atomic-level framework of the exciton condensation in TiSe$_2$. Our lr-TDDFT approach both accurately reproduces the experimental spectra as well as explains the complex interplay among the thermal field and CDW order effects that were recently observed in exciton condensation experiments. We find that the soft mode that characterizes the exciton condensation can be attributed to interband electronic transitions. Furthermore, the fragile temperature dependence of the exciton condensation is the combined effect of the CDW and thermal carriers: Below $T_c$, the periodic lattice distortions hamper the spontaneous formation of the exciton by introducing a CDW gap; above $T_c$, a higher electronic temperature produces sufficient thermal carriers and introduces a dissipation channel. This explains why the soft mode is only observed at $T_c$. As such, our \textit{ab initio} framework provides critical mechanistic insight into recent exciton condensation experiments and presents additional avenues to experimentalists for further manipulating exciton condensates in TiSe$_2$.

\section{Acknowledgement}
C. L. acknowledges the helpful discussions with Dr. Sangeeta Sharma. This work was supported by the U.S. Department of Energy, Office of Science, Early Career Research Program under Award No. DE‐SC0016269.

\end{document}